\documentclass[mypaper,8pt,twoside]{CoAst}
\usepackage{epsf,graphicx,fancyhdr}
\usepackage{sfmath}

\renewcommand{\chaptermark}[1]{\markboth{#1}{}}

\newcommand{\kopf}{\small\itshape Comm. in Asteroseismology\\ Vol. 150, 2007}
\newcommand{\Authors}[1]{\begin{center}\normalsize\bf\sf #1 \end{center}}

\renewcommand{\author}[1]{\begin{center}\normalsize\bf\sf #1 \end{center}}
\newcommand{\Address}[1]{\begin{center}\small\sf #1 \end{center}}

\newcommand{\References}[1]{\vspace{2.4mm}\begin{flushleft}{\large References\\}\vspace*{1mm}\small #1 \end{flushleft}}

\newcommand{\chapterDSSN}[2]{\chapter[\sf\normalsize #1\\ \footnotesize \hspace*{5mm}by #2 \sf\normalsize][]{#1\\}\rhead[\fancyplain{}{\sf\footnotesize \center{#1}}]{\fancyplain{}{\sffamily\thepage}}\lhead[\fancyplain{\kopf}{\sffamily\thepage}]{\fancyplain{\kopf}{\sf\footnotesize \center{#2}}}}

\newcommand{\figureDSSN}[5]{\begin{figure}[#4]
\centering
\includegraphics*[#5]{#1}
\caption{#2}
\label{#3}
\end{figure}}

\renewcommand{\chaptername}{}
\newcommand{\acknowledgments}[1]{\vspace*{5mm}\noindent\begin{bf}Acknowledgments. \end{bf} #1}

\def\rfr{\smallskip\par\noindent
        \hangindent=7truemm
        \hangafter=1}

\begin{document}
\sf

\chapterDSSN{Solar-like oscillations in open cluster stars}{D. Stello et al.}

\Authors{D. Stello,$^{1}$ H. Bruntt,$^{1}$ T. Arentoft,$^{2,3}$
R. L. Gilliland,$^{4}$ J. Nuspl,$^{5}$ S.-L. Kim,$^{6}$ Y. B. Kang,$^{6}$
J.-R.~Koo,$^{6}$ J.-A.~Lee,$^{6}$ C.-U. Lee,$^{6}$ C. Sterken,$^{7}$
A. P. Jacob,$^{1}$ S.~Frandsen,$^{2,3}$ Z.\ E. Dind,$^{1}$ 
H. R. Jensen,$^{2}$ R.~Szab\'o,$^{5}$ Z. Csubry,$^{5}$ L.\ L.~Kiss,$^{1}$
M. Y. Bouzid,$^{7}$ T.\ H.~Dall,$^{8}$ \\ T. R. Bedding,$^{1}$
H. Kjeldsen\,$^{2,3}$}
\Address{$^{1}$\,School of Physics, University of Sydney, NSW 2006, 
Australia\\
$^{2}$\,Institut for Fysik og Astronomi (IFA), Aarhus Universitet,
  8000 Aarhus, Denmark\\
$^{3}$\,Danish AsteroSeismology Centre, Aarhus Universitet, 8000 Aarhus, 
Denmark\\
$^{4}$\,Space Telescope Science Institute, 3700 San Martin Dr., Baltimore, 
USA\\
$^{5}$\,Konkoly Observatory, 1525 Budapest, PO Box 67, Hungary\\
$^{6}$\,Korea Astronomy and Space Science Institute, Daejeon 305-348, 
Korea\\
$^{7}$\,Vrije Universiteit Brussel, Pleinlaan 2, B-1050 Brussels, 
Belgium\\
$^{8}$\,European Southern Observatory, Casilla 19001, Santiago 19, Chile}

\section{Introduction} 
Asteroseismology of stellar clusters is potentially
a powerful tool. The assumption of a common age, distance, and chemical
composition provides constraints on each cluster member, which
significantly improves the asteroseismic output. Hence, detecting
oscillations in cluster stars in a range of evolutionary stages holds
promise of providing more stringent tests of stellar evolution
theory. Driven by this great potential, we carried out multi-site
observations aimed at detecting solar-like oscillations in the red giant
stars in the open cluster M67 (NGC 2682). To obtain these data we observed 
for 43 days with nine 0.6-m to 2.1-m class telescopes in
January and February 2004 (Stello et al.\ 2006). The photometric time
series comprises roughly 18000 data points. The properties of the red giant
stars we present here are given in Table~\ref{stellotable1} and their
location in the colour-magnitude diagram is shown in
{Fig.~\ref{stellofig1} (left panel)}. 

\begin{table}
%\begin{table*}
\begin{center}
\caption{Properties of red giant target stars. Luminosities and
   temperatures are from photometry (Montgomery et al.\ 1993). Estimates
    of oscillation amplitudes, characteristic frequencies and large
    separations are from known scaling relations in the literature
    (Kjeldsen \& Bedding 1995, Brown et al.\ 1991), Cross references are to
    Sanders (1977) and Gilliland et al.\ (1991).}   
%\scriptsize
\begin{tabular}{@{}rrrrrrr@{/}l@{}}
\noalign{\smallskip}
\hline  
\hline  
\noalign{\smallskip}
ID & $L/L_{\odot}$ & $T_{\mathrm{eff}}$ & $\delta L/L$ & 
  $\nu_{\mathrm{max}}$ & $\Delta\nu_{0}$ & \multicolumn{2}{c}{Cross-ref.} \\  
   &               & K                  & $\mu$mag     & 
  $\mu$Hz              & $\mu$Hz         & \multicolumn{2}{c}{}           \\ 
\noalign{\smallskip}
\hline  
\hline 
\noalign{\smallskip}
% 3&  250.6& 3920& 2080&   3.7&  0.7&   S978&G8  \\ 
%11&  243.5& 3960& 1980&   3.9&  0.8&  S1250&G4  \\
% 4&   87.0& 4330&  592&  15.1&  2.2&  S1016&G53 \\
%21&   51.9& 4727&  296&  34.4&  4.2&  S1592&--   \\
 8&   50.8& 4750&  287&  35.8&  4.3&  S1010&G2  \\
 9&   50.2& 4772&  281&  36.8&  4.4&  S1084&--  \\
10&   48.2& 4727&  275&  37.0&  4.4&  S1279&G7  \\
%20&   47.1& 4772&  264&  39.2&  4.6&  S1479&--  \\
 2&   46.4& 4727&  265&  38.4&  4.6&  S1074&--  \\
18&   45.8& 4772&  256&  40.3&  4.7&  S1316&--  \\
%16&   40.3& 4703&  232&  43.5&  5.0&  S1221&--  \\
\hline  
\noalign{\smallskip}
 5&   25.4& 4815&  140&  74.8&  7.6&  S1054&G9  \\
17&   22.4& 4835&  122&  86.0&  8.4&  S1288&--  \\
 7&   20.2& 4854&  109&  96.9&  9.2&   S989&G12 \\
%19&   18.7& 4873&  101& 106.0&  9.9&  S1254&--  \\
15&   16.9& 4873&   91& 117.3& 10.6&  S1277&--  \\
\hline  
\noalign{\smallskip}
14&   11.2& 4945&   58& 187.3& 15.2&  S1293&--  \\
%12&   11.0& 4947&   57& 190.2& 15.4& S1264a&G15 \\
13&    9.9& 4966&   51& 213.2& 16.8&  S1305&--  \\
% 6&    9.2& 4971&   48& 230.2& 17.8&  S1103&--  \\
\hline  
\hline
\label{stellotable1}
\end{tabular}
\end{center}
%\end{table*}
\end{table}

\figureDSSN{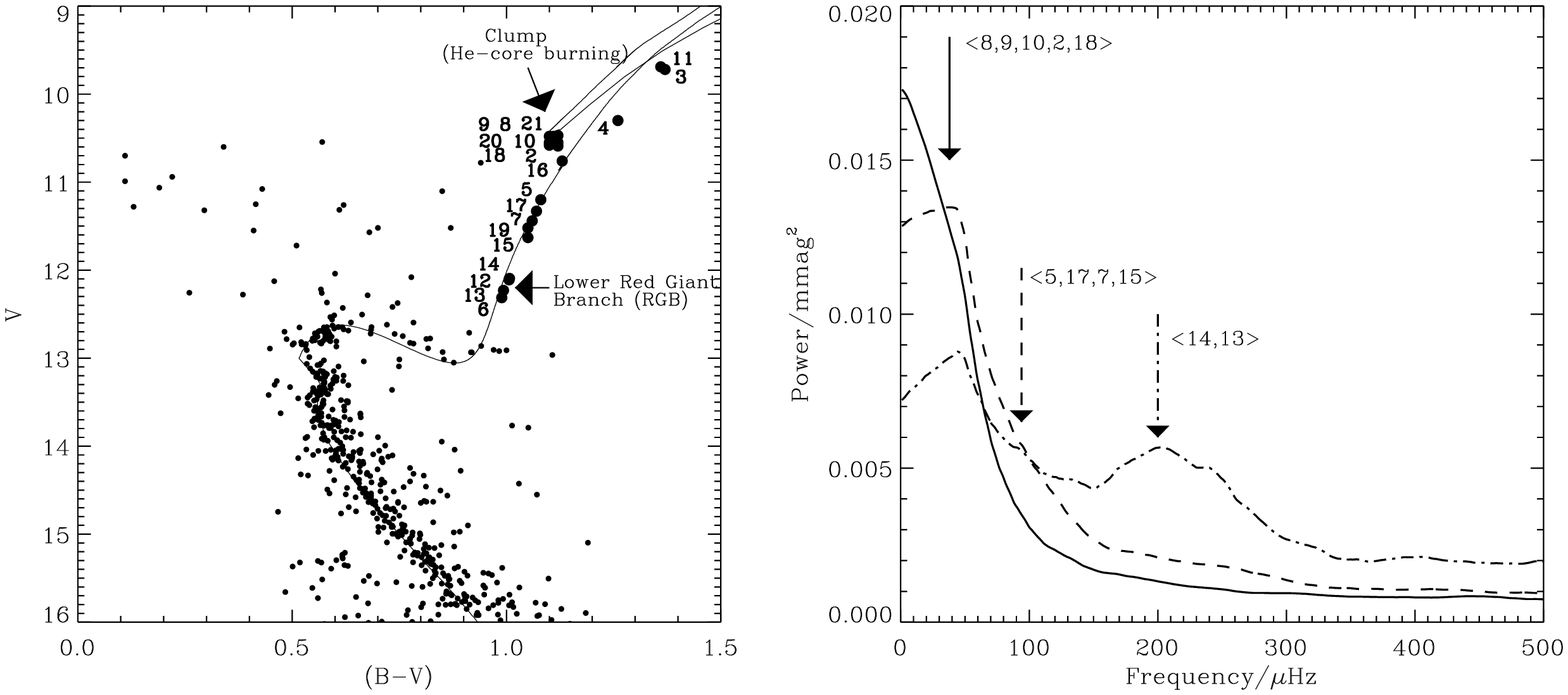}{Left Panel: The colour-magnitude diagram of
the open cluster M67. The red giant target stars and their ID are
indicated. Right Panel: Average power distributions for three groups of
stars sorted according to luminosity: most luminous (clump stars),
intermediate, and least luminous (lower RGB). Arrows show expected
locations of solar-like oscillations (see Table~\ref{stellotable1}). Only
stars with a white-noise level lower than 50$\,\mu$mag have been
used.}{stellofig1}{!ht}{clip,angle=0,width=115mm}

\section{Results}

Mean levels in the Fourier spectra in the frequency interval
300--900$\,\mu$Hz, corresponding to white noise, reach $20\,\mu$mag for
the stars with the lowest noise. In many stars we see apparently high
levels of non-white noise, but the detailed temporal variation of the
noise is unknown. We are therefore not able to clearly disentangle the
noise and stellar signal in the analysis. However, we do see evidence of
excess power in the Fourier spectra, shifting to lower frequencies for
more luminous stars, consistent with expectations (Fig.~\ref{stellofig1};
right panel).  If the observed power excesses were due to stellar
oscillations, this result would show great prospects for asteroseismology
in stellar clusters. A more detailed analysis will be given by Stello et
al.\ (2007).

\acknowledgments{This paper has been supported by the Astronomical Society
of Australia.}

\References{
\rfr Brown T.\ M., Gilliland R.\ L., Noyes R.\ W., Ramsey L.\ W., 1991, ApJ 368, 599
\rfr Gilliland R.\ L., Brown T.\ M., Thomson D.\ T., Schild R.\ E., Jeffrey W.\ A., Penprase B.\ E., 1991, AJ 101, 541
\rfr Kjeldsen H., Bedding T.\ R., 1995, A\&A 293, 87
\rfr Montgomery K.\ A., Marschall L.\ A., Janes K.\ A., 1993, AJ 106, 181
\rfr Sanders W.\ L., 1977, A\&AS 27, 89
\rfr Stello D., et al., 2006, MNRAS, 373, 1141
\rfr Stello D., et al., 2007, MNRAS, accepted}

\end{document}